# Fault Localization for Java Programs Using Probabilistic Program Dependence Graph

A.Askarunisa[1], T. Manju[2] and B. Giri Babu

[1] **Computer Science and Engineering, Thiagarajar College Of Engineering,**
**Madurai, Tamilnadu, India**
*aacse@tce.edu*

[2] **Computer Science and Engineering, Thiagarajar College Of Engineering,**
**Madurai, Tamilnadu, India**
*manju.dts27@gmail.com*

[3]**Honeywell Technology Solutions Lab. Pvt. Ltd,**
**Madurai, Tamilnadu, India**
*giri1209@gmail.com*

**Abstract**
Fault localization is a process to find the location of faults. It determines the root cause of the failure. It identifies the causes of abnormal behaviour of a faulty program. It identifies exactly where the bugs are. Existing fault localization techniques are Slice based technique, Program-Spectrum based Technique, Statistics Based Technique, Program State Based Technique, Machine learning based Technique and Similarity Based Technique. In the proposed method Model Based Fault Localization Technique is used, which is called *Probabilistic Program Dependence Graph* . Probabilistic Program Dependence Graph (PPDG) is an innovative model that scans the internal behaviour of the project. PPDG construction is enhanced by Program Dependence Graph (PDG). PDG is achieved by the Control Flow Graph (CFG). The PPDG construction augments the structural dependences represented by a program dependence graph with estimates of statistical dependences between node states, which are computed from the test set. The PPDG is based on the established framework of probabilistic graphical models. This work presents algorithms for constructing PPDGs and applying fault localization.

***Keywords:*** *Probabilistic Program Dependence Graph (PPDG), Fault Localization, Program Dependence Graph (PDG).*

## 1. Introduction

In the software industry, developers usually rely on testing to confirm that changes to the software achieve their intentions and do not introduce unexpected side effects. Typically, testing involves executing a large number of test cases and thus is very time-consuming. For instance, the industrial collaboration of Elbaum et al. [12,14] reported that it costs seven weeks to execute the entire test suite of one of their products.

To cope with the preceding situation, researchers have proposed various techniques for fault localization [3,5,9,10] to find the faults exactly where is. A fault is nothing but the bugs. It is always challenging for programmers to effectively and efficiently remove bugs. Furthermore, to debug, programmers must first be able to identify exactly where the bugs are, which is known as *fault localization.*

Fault Localization is defined as the process of finding the faults of any program. There are various techniques for fault localization. Section 2 defines various existing techniques. Section 3 explains about fault localization using Statistical Bug Isolation (SBI). Section 4 defines our approach. Section 5 gives the experimental results. Section 6 defines the performance results between SBI and PPDG and Section 7 concludes.

## 2. Related Work

Fault Localization is an intensively studied research topic in regression testing. Techniques for fault localization aim to improve the rate of fault detection. In the literature, there are several lines of research on fault localization. The first line of research is to study techniques for fault localization.

First technique is program slicing. It is a commonly used technique for debugging. Reduction of the debugging search domain via slicing is based on the idea that if a test case fails due to an incorrect variable value at a statement, then the bug should be found in the static slice associated with that variable-statement pair [22]. Lyle & Weiser extended the above approach by constructing a program dice to further reduce the search domain for possible locations of a fault [15]. A disadvantage of this technique is that it might generate a dice with certain statements which should not be included. Studies such as [2], [19], [31] use the dynamic slicing concept to program debugging. An alternative is to use execution slicing and dicing to locate program bugs [24], where an execution slice with respect to a given test case contains the set of code executed by this test.

Second technique is program-spectrum based technique. A program spectrum records the execution information of a program. When the execution fails, such information can be used to identify suspicious code that is responsible for the failure. Tarantula [10] is a popular fault localization technique based on the *executable statement hit* spectrum. It uses the execution trace information in terms of how each test covers the executable statements, and the corresponding execution result (success or failure) to compute the suspiciousness of each statement. One problem with Tarantula is that it does not distinguish the contribution of one failed test case from another, or one successful test case from another. To overcome this problem, Wong et al. [22] propose that, with respect to a piece of code, the contribution of the $n$th failed test in computing its suspiciousness is larger than or equal to that of the $(n+1)$th failed test. Renieris & Reiss [18] propose a program spectrum-based technique such as nearest neighbor, which contrasts a failed test with another successful test that is most similar to the failed one in terms of the "distance" between them. If a bug is in the difference set between the failed execution and its most similar successful execution, it is located. For a bug that is not contained in the difference set, the technique continues by first constructing a program dependence graph, and then including and checking adjacent un-checked nodes in the graph step by step until the bug is located.

Third is statistics based techniques. Several statistical fault localization techniques have also been proposed, such as Liblit05 [11], and SOBER [12], which rely on the instrumentations and evaluations of predicates in programs to produce a ranking of suspicious predicates, which can be examined to find faults. They are also limited to bugs located in predicates, and offer no way to attribute a suspiciousness value to all executable statements. Wong et al. propose a cross tabulation (crosstab) based statistical technique which uses only the coverage information of each executable statement, and the execution result with respect to each test case. It does not restrict itself to faults located only in predicates. More precisely, a crosstab is constructed for each statement with two column-wise categorical variables of "covered," and "not covered;" and two row-wise categorical variables of "successful execution," and "failed execution".

Fourth is Program state based technique. A program state consists of variables, and their values at a particular point during the execution. A general approach for using program states in fault localization is to modify the values of some variables to determine which one is the cause of erroneous program execution. Zeller, et al. propose a program state-based debugging approach, delta debugging [5], to reduce the causes of failures to a small set of variables by contrasting program states between executions of a successful test and a failed test via their memory graphs. Based on delta debugging, Cleve & Zeller [6] propose the cause transition technique to identify the locations and times where the cause of failure changes from one variable to another. A potential problem is that the cost is relatively high. Another problem is that the identified locations may not be where the bugs reside. Gupta et al. [8] try to overcome these issues by introducing the concept of failure inducing chops. Predicate switching proposed by Zhang, et al. is another program state-based fault localization technique where program states are changed to forcefully alter the executed branches in a failed execution. A predicate whose switch can make the program execute successfully is labeled as a critical predicate. Wang & Roychoudhury [20] present a technique that automatically analyzes the execution path of a failed test, and alters the outcome of branches in that path to produce a successful execution. The branch statements whose outcomes have been changed are recorded as bugs.

Fifth is Machine learning based technique. Machine learning techniques are adaptive, and robust; and have the ability to produce models based on data, with limited human interaction. The problem at hand can be expressed as trying to learn or deduce the location of a fault based on input data such as statement coverage, etc. Wong et al. [21] propose a fault localization technique based on a back-propagation (BP) neural network, which is one of the most popular neural network models in practice. The statement coverage of each test case, and the corresponding execution result, are used to train a BP neural network. Then, the coverage of a set of *virtual* test cases that each covers only one statement in the program are input to the trained BP network, and the outputs can be regarded as the likelihood of the statements being faulty. Briand et al. [4] use the C4.5 decision tree algorithm to construct a set of rules that might classify test cases into various partitions such that failed test cases in the same partition most likely fail due to the same fault.

Other techniques involved are data mining-based (e.g., Cellier et al. [5] which discuss a combination of association rules and Formal Concept Analysis (FCA) to assist in fault localization), and model-based (e.g., [10]). Similarity-based coefficients such as Ochiai & Jaccard.

In our paper, we study the problem of model based fault localization. Our work differs from previous research on fault localization as follows. First, the use of PPDG for fault localization. Second, this paper presents the first empirical comparison of techniques.

Our approach involves the fault localization for java programs using PPDG.

## 3. Fault Localization Using SBI

Liblit et al. [21] propose Statistical Bug Isolation (SBI) for computing the suspiciousness of a predicate *P* in a program, thus:

$$Failure(P) = \frac{failed(P)}{passed(P) + failed(P)}$$

The function *failed* (*passed*, respectively) tallies the number of test cases for which *P* is evaluated to be false (true). For ease of comparison with other fault localization techniques, Yu et al. [29] adapt the equation to calculate the suspiciousness of a statement *s* as follows:

$$suspiciousnesss(s) = \frac{failed(s)}{passed(s) + failed(s)}$$

The function *failed* (*passed*, respectively) tallies the number of test cases for which *s* is evaluated to be false (true).

## 4. Our approach

In this paper, we study the problem of fault localization using model based technique. Our research differs from previous research on fault localization as follows.
- The use of Probabilistic Program Graph for fault localization.
- This work presents the how the fault localization takes place in java programs using this technique.

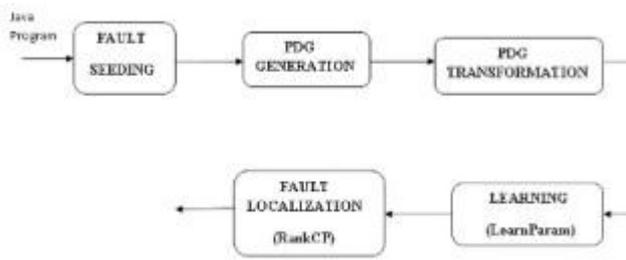

Fig. 1 Overall architecture for fault localization

Our approach involves the PPDG generation and fault localization. PPDG is an innovative model of a program's internal behaviour over a set of test inputs. It facilitates probabilistic analysis and reasoning about uncertain program behaviour, particularly those associated with faults. The PPDG is based on the established framework of probabilistic graphical models. It scans each and every state nodes for fault in a program. Since it is a graphical representation testers can easily find exactly where the fault is. PPDG is nothing but transformed PDG. This transformation is achieved by learning. For learning LearnParam algorithm is used. This will transform the predicate nodes and self loop nodes by adding additional node as its parent.

LearnParam algorithm is used to generate the PPDG. Fault localization is done by RankCP algorithm. It is used to find the probabilistic distribution of each node. It ranks each node using probability and the node having less probability is considered to be most suspicious. In LearnParam algorithm the execution trace is taken as input and evaluates the probability for each node based on the dependences to the node.

4.1 Decision Variables

A probabilistic graphical model is an annotated graph that captures the probabilistic relationships among a set of random variables. The nodes in the graph represent random variables and the edges represent conditional dependences between the random variables.

4.2 Objective Function

To achieve the goal, we have to find the probability for each node. It is categorized as

4.2.1 Node with no parents

For a node with no parents, our technique estimates the probabilities ($p(X_j = x_{ji})$) of the nodes as given in equation (1)

$$p(X_j = x_{ji}) = \frac{n(X_j = x_{ji})}{n(X_j)} \quad \text{------(1)}$$

where $n(X_j = x_{ji})$ is the number of times node ($X_j$) is in state $x_{ji}$ across all node-state traces and $n(X_j)$ is the number of times the node $X_j$ occurs across all node-state trace.

4.2.1 Node with parents

For a node with parents, our technique estimates the probabilities ($p(X_j = x_{ji}|Pa(X_j) = pa_{ji})$) ) of the node as given in equation (2) below.

$$P(X_j = x_{ji} \mid Pa(X_j = pa_{ji})) = \frac{n(X_j = x_{ji}, Pa(X_j = pa_{ji}))}{n(Pa(X_j = pa_{ji}))} \quad \text{--- (2)}$$

where $n(X_j=x_{ji}, pa(X_j=pa_{ji}))$ is the number of times node Xj and its parents assume a specific state configuration across all node-state traces and $n(pa(X_j=pa_{ji}))$ is the number of times $pa(X_j=pa_{ji})$ across all node-state traces. A state configuration is a set of states assigned to a set of nodes in the PPDG.

### 4.3 Constraint System

The constraint for finding the fault is to find the conditional probability for each node. The node having less probability is considered to be most suspicious node and deemed to be a fault node. It is found by using RankCP algorithm.

### 4.4 Fault Seeding

To evaluate the performance of fault localization, we require the following approaches of faults seeding that are classified

1. Mutation
2. Hand seeding.

#### 4.4.1 Mutation

Faults are inserted that are as realistic as possible and that involved code deleted from, inserted into, or modified in the versions. The following lists of types of faults are considered.
- Faults associated with variables, such as with definitions of variables, redefinitions of variables, deletions of variables, or changes in values of variables in assignment statements;
- Faults associated with control flow, such as addition of new blocks of code, deletions of paths, redefinitions of execution conditions, removal of blocks, changes in order of execution, new calls to external functions, removal of calls to external functions, addition of functions, or deletions of functions;
- Faults associated with memory allocation, such as not freeing allocated memory, failing to initialize memory, or creating erroneous pointers.

#### 4.4.2 Hand Seeding

Faults are inserted that are as realistic as possible and that involved code deleted from, inserted into, or modified in the versions. The following lists of types of faults are considered.
- Faults associated with variables, such as with definitions of variables, redefinitions of variables, deletions of variables, or changes in values of variables in assignment statements;
- Faults associated with control flow, such as addition of new blocks of code, deletions of paths, redefinitions of execution conditions, removal of blocks, changes in order of execution, new calls to external functions, removal of calls to external functions, addition of functions, or deletions of functions;
- Faults associated with memory allocation, such as not freeing allocated memory, failing to initialize memory, or creating erroneous pointers.

The first approach would allow to generate a large number of faults. The second approach cannot cost-effectively produce a large number of faults. Thus, we chose the first approach

### 4.5 PDG Generation

PDG is generated for a given java program which is to be tested. It is the combination of both control flow graph and data flow graph. Using the control flow graph, we can informally define both control dependence and data dependence. In a control flow graph G, node n1 is control dependent on node n2 if n2 has outgoing edges e1 and e2 such that 1) every path in G starting with e1 and ending with an exit node contains n1 and 2) there is a path starting with e2 and ending with an exit node that does not contain n1. A probabilistic graphical model is an annotated graph that captures the probabilistic relationships among a set of random variables. The nodes in the graph represent random variables and the edges represent conditional dependences between the random variables. The nodes in the PDG are labelled with the line numbers of the corresponding statements in the program. Solid edges represent control dependences between nodes and dotted edges represent data dependences between nodes. Labels on the control dependence edges are either "T" for true or "F" for false. Labels on the data dependence edges represent the variables involved in the data flows between the nodes.

```
    ┌ import java.io.*;
  1 │ public class LineNumberReaderExample{
    │ public static void main(String[] args) throws Exception{
    └ File file = null;

  2   FileReader freader = null;
  3   LineNumberReader lnreader = null;

    ┌ try{
  4 │ file = new File("dgraph.java");
  5   freader = new FileReader(file);
  6   lnreader = new LineNumberReader(freader);
  7   String line = "";
  8   while ((line = lnreader.readLine()) != null){
  9   System.out.println("Line:   " + lnreader.getLineNumber() + ": " +
      line);
    
    
    ┌ finally{
    │ freader.close();
 10 │ lnreader.close();
```

Fig. 2 Sample java Program

To illustrate the above considered an example program that finds the line number of a java program, shown in Fig. 2. Fig. 3 shows the control flow graph (CFG). In the graph, each node is labelled with the number of the program statement that it represents, and each edge shows the flow of control between the corresponding statements.

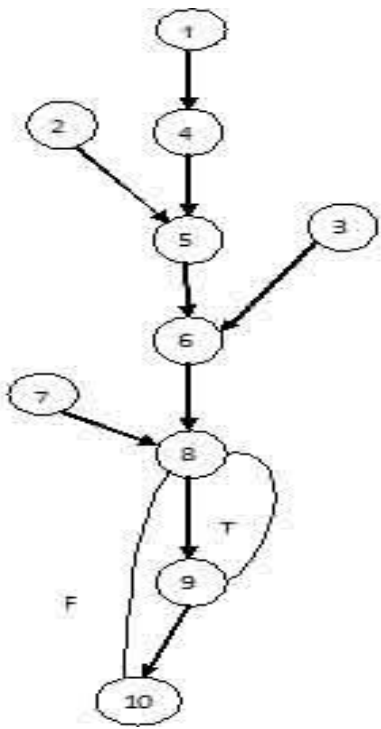

Fig. 3 Control Flow Graph (CFG)

For example, node 1 represents the first statement in the program and node 10 represents the last statement in the program. For another example, node 8 has two outgoing edges: Edge (8,9) is taken if the condition at 8 is true (i.e., the while loop is entered) and edge (8, 10) is taken if the condition at 8 is false. Using the control flow graph, we can informally define both control dependence and data dependence.

From the control flow graph generated, the PDG is generated as shown in Fig. 4

PDG is generated with the combination of control flow graph and data flow graph. In this PDG, the control flow is derived from the previous step. The data flow, for example line 3 has a variable lnreader. It flows across line 6,8 and 9 simultaneously. Thus PDG includes both data flow and control flow.

4.6 PDG Transformation

PDG transformation it by structurally changing the PDG and specifying states at nodes in the PDG, which results in a transformed PDG. It structurally transforms the PDG by adding nodes and edges to the predicate nodes and self loop nodes. It is because while calculating the conditional probability distribution of each node the dependences between the nodes get duplicated.

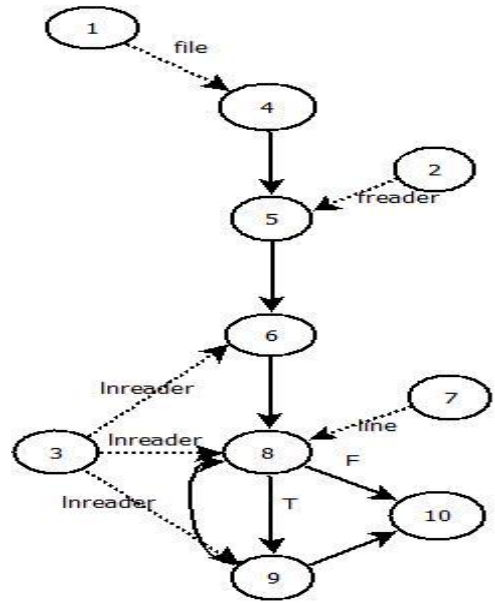

Fig. 4 Program Dependence Graph (PDG).

During this step, our technique
1) structurally transforms the PDG by adding nodes and edges to it and
2) specifies the states of the nodes.

We call the graph that results after transforming the PDG the transformed PDG. The technique assigns to each node in a program's transformed PDG a finite set of discrete abstract states, each of which represents a set of related concrete states of the corresponding statement. Hereafter, we use the term "state" to refer to an abstract state.

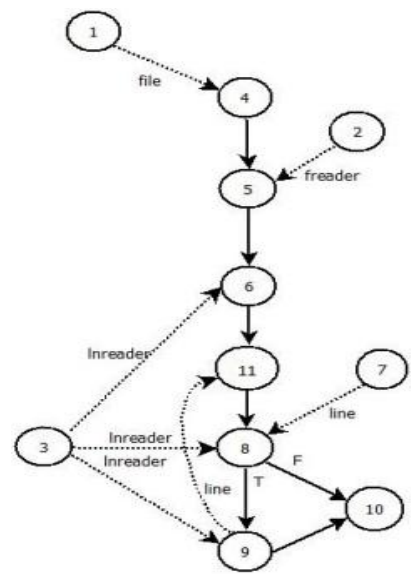

Fig. 5 Transformed PDG (PPDG)

The states of a node must be mutually exclusive. i.e., a node cannot be in two different states at the same time. The state of a PPDG node abstracts a part of the program's state that pertains to the node when the program executes. There are different ways to model this "local" concrete state. In this work, we model it in one or both of two ways depending on whether the node represents a branch predicate, a statement that uses one or more variables, or both. These characterizations are intended to reflect certain aspects of a node's concrete state that are relevant to applications, such as fault localization.

Generally, the transformation is enhanced by adding new nodes and edges to the predicate nodes and self loop nodes. In the above example there is a predicate node and not self loop node. Node 8 is the predicate node( while loop) . So new node 11 is added to it and edge labelled line is added to the node11.

### 4.7 Learning

Learning estimates the parameters of the PPDG from the set of execution data generated by executing the instrumented program with its test suite. Different kinds of execution data (e.g., coverage or trace information) might be used to estimate the parameters of the PPDG. In this work, our technique uses node-state traces. A node-state trace is a sequence of executed nodes, along with their active states, in the transformed PDG. Node-state traces is to estimate the parameters of the PPDG. Each $D_k £ D$ is a node state trace. A node can appear multiple times in the trace, and the states that the node assumes can be different. In this work, we present a batch-learning algorithm called LearnParam.

```
Algorithm: LearnParam
Input: D = {D_k}_{k=1}^n; transformed PDG
Output: PPDG
1 foreach D_k ∈ D do
2     for j = 1 to Length(D_k) do
3         if Pa(X_j) = ∅ then
4             increment n(X_j = x_ji) by 1, where x_ji is the
              current state of X_j
5         else
6             increment n(X_j = x_ji, Pa(X_j) = pa_ji) by 1, where
              x_ji is the current state of X_j and where pa_ji
              represents the current state configuration of the
              parents of X_j
7         end
8     end
9 end
10 compute probabilities of X_j using equations (5), (2) return PPDG
```

Fig. 6 LearnParam Algorithm

LearnParam algorithm is used to estimate the parameters. Learning the parameters of the PPDG consists of estimating conditional probability distributions, which are represented as tables called conditional probability tables (CPTs), because the states of the nodes in the transformed PDG are discrete. It gets each data trace as input and calculates the probability for each node. Different kinds of execution data might be used to estimate the parameters of the PPDG. The output will be PPDG.

For the above example, using the LearnParam algorithm probability is calculated. The probability for each node is calculated and given in the Table 1.

Table 1: Conditional Probability Calculation

| NODE | PROBABILITY |
|------|-------------|
| 1 | 1 |
| 2 | 1 |
| 3 | 1 |
| 4 | 1 |
| 5 | 0.6 |
| 6 | 1.0 |
| 7 | 1.0 |
| 8 | 0.5 |
| 9 | 0.66 |
| 10 | 0.4 |
| 11 | 0.5 |

### 4.8 Fault Localization

The performance of fault localization is encountered by the RankCP algorithm.

```
Algorithm: RankCP
Input: node-state trace:{X_j : x_ji}_{j=1}^n; PPDG
Output: ranked nodes with state configurations
1 for j = 1 to n do
2     prob ← p(X_j = x_ji | Pa(X_j) = pa_ji)
3     if prob < lowest_prob(X_j) then
4         lowest_prob(X_j) ← prob
5         index(X_j) ← j
6         configuration(X_j) ← {x_ji ∪ pa_ji}
7     end
8 end
9 rank nodes in ascending order by probability, break ties using
  indexes
10 return ranked nodes with state configurations
```

Fig. 7 RankCP Algorithm

RankCP algorithm analyzes a single failed execution at a time, and ranks nodes in the PPDG. RankCP ranks nodes based on the conditional probabilities of nodes given the states of their parent nodes which reflect how the parents influence their children. Our hypothesis is that RankCP will often detect the first place in a failing execution, where a node (Xj) assumes an unusual state, given the states of its parents, thus indicating a possible cause of the failure. RankCP ranks a node Xj that a state whose probability is low, given the states of

Xj's parents, as highly suspicious. Our choice of this conditional probability as an inverse measure of suspiciousness is based on preliminary studies we conducted that showed that faults tend to be associated with low probability nodes.

For a given program, RankCP inputs its PPDG and a node-state trace generated by a failing execution, and it returns a list of nodes ranked from most suspicious to least suspicious. Each node is also associated with a node-parent state configuration. RankCP processes a trace from beginning to end. As it processes the trace, it computes the conditional probability of a node's current state ($x_{ji}$) given the current state configuration ($pa_{ji}$) of its parents. Then, RankCP records for each node the lowest value lowest_prob of this probability (lines 3 and 4). RankCP also keeps track of the index of a node in the trace in the index variable (line 5). RankCP associates a node-parent state configuration with a node using the configuration variable (line 6). After RankCP has processed the trace, it ranks the nodes by their lowest_prob values, and if two nodes have the same lowest_prob values, the algorithm ranks the node with the lower index value higher (line 9). The algorithm returns the ranked nodes with their associated state configurations In the above example, node 10 has the lowest probability 0.4. So it is considered to be the fault node.

## 5. Experiments

In our experiments, we study the effectiveness of the reduction strategies by evaluating their fault detection rate. A program under test can be assessed by counting and classifying the discovered faults.

*Subject application and Test suites :* We used 21 Java programs and have generated test cases by calculating the cyclomatic complexity, which gives the upper bound for the maximum number of test cases. The details of the various application programs and their corresponding metrices are shown in Table 3. The programs Aes, Fiestel, Playfair, Sdes, Trans, Des, Hill cipher, Rc4, Mono alphabetic substitution, Caesar cipher, Diffie Hell man are all programs related to network security algorithms. All the other programs are simple programs done by our students. We have designed a tool to calculate the various OO metrics namely Lines of Code (LOC), Weighted methods per class (WMC), Depth of Inheritance tree (DIT), Coupling between object classes (CBO), Response for a class (RFC), Lack of cohesion in methods (LCOM), Total lines of code (TLOC), Executable Lines of Code (ELOC), Number of operands (OPn), Number of Operators (OPr) etc. These metrics help the test manager to determine the quality of the programs.

*Evaluation Metrics:* For evaluating the reduction techniques, we have injected hand seeded faults into our programs. We have included faults like arithmetic operator faults, logical operator faults and relational faults. In our example we injected 13 faults and identified which test case identifies which faults as shown in Table 5.

From the result we have noticed that our proposed work has less faults than the previous technique Statistical Bug Isolation(SBI)[3]. The Comparative table is shown in Table.5

Table2: EvaluationMeasures

| Metrics / Programs | ELOC | OPr | OPn | Exec time (ET) | WBC | DIT | TLOC | CBO | RFC | LCOM |
|---|---|---|---|---|---|---|---|---|---|---|
| Aes | 631 | 1461 | 2331 | 18.21437 | 11 | 1 | 647 | 9 | 12 | 45 |
| Fiestel | 163 | 372 | 372 | 8.950757 | 6 | 1 | 203 | 5 | 11 | 10 |
| Playfair | 82 | 306 | 426 | 8.814598 | 9 | 1 | 173 | 7 | 18 | 28 |
| Sdes | 263 | 561 | 1055 | 9.00457 | 12 | 1 | 295 | 5 | 7 | 55 |
| Trans | 64 | 199 | 280 | 12.62874 | 3 | 1 | 129 | 11 | 14 | 1 |
| Des | 159 | 417 | 822 | 36.569842 | 8 | 1 | 214 | 10 | 12 | 21 |
| Hill cipher | 197 | 365 | 541 | 11.254786 | 7 | 1 | 197 | 12 | 17 | 15 |
| Rc4 | 120 | 294 | 454 | 5.45267 | 5 | 1 | 207 | 6 | 8 | 6 |
| Diffie hell man | 78 | 132 | 222 | 4.78965 | 3 | 1 | 156 | 4 | 7 | 6 |
| Mono alphabetic substitution | 346 | 243 | 267 | 5.60986 | 6 | 1 | 459 | 6 | 7 | 5 |
| Rail fence | 123 | 69 | 54 | 4.786650 | 3 | 1 | 169 | 5 | 3 | 4 |
| Caesar cipher | 238 | 53 | 48 | 4.09863 | 5 | 1 | 287 | 5 | 2 | 3 |
| Bubble sort | 29 | 69 | 141 | 0.0065487 | 2 | 2 | 63 | 6 | 4 | 0 |
| Game Play | 34 | 17 | 20 | 0.054543 | 2 | 1 | 54 | 2 | 2 | 2 |
| Sample | 50 | 14 | 16 | 0.085432 | 2 | 1 | 64 | 2 | 3 | 1 |
| Salary Calculation | 80 | 26 | 29 | 0.1596898 | 4 | 1 | 43 | 1 | 2 | 2 |
| Discount calculation | 45 | 9 | 12 | 0.06598 | 3 | 1 | 59 | 1 | 2 | 2 |
| Type of a triangle | 37 | 7 | 9 | 0.000345 | 1 | 1 | 45 | 1 | 1 | 1 |
| Link vector | 34 | 9 | 6 | 0.035198 | 2 | 1 | 53 | 2 | 1 | 2 |
| Calculating Grade | 26 | 6 | 5 | 0.00285 | 1 | 1 | 33 | 2 | 2 | 2 |
| Distance vector | 39 | 11 | 10 | 0.08345 | 3 | 1 | 51 | 2 | 2 | 2 |

Table 3: Performance Analysis

| NAME OF PROGRAM | NUMBER OF FAULTS | FAULTS IDENTIFIED | |
|---|---|---|---|
| | | SBI | PPDG |
| Aes | 33 | 27 | 30 |
| Fiestel | 27 | 21 | 25 |
| Playfair | 17 | 12 | 15 |
| Sdes | 19 | 10 | 15 |
| Trans | 15 | 10 | 12 |
| Des | 47 | 40 | 43 |
| Hill cipher | 31 | 25 | 28 |
| Rc4 | 43 | 35 | 39 |
| Diffie hell man | 23 | 18 | 21 |
| Mono alphabetic substitution | 15 | 10 | 13 |
| Rail fence | 17 | 12 | 15 |
| Caesar cipher | 18 | 13 | 15 |
| Bubble sort | 19 | 10 | 15 |
| Game Play | 9 | 5 | 7 |
| Sample | 5 | 3 | 4 |
| Salary Calculation | 15 | 10 | 13 |
| Discount calculation | 10 | 6 | 8 |
| Type of a triangle | 25 | 20 | 23 |
| Link vector | 45 | 35 | 40 |
| Calculating | 20 | 14 | 17 |
| Distance vector | 35 | 29 | 32 |

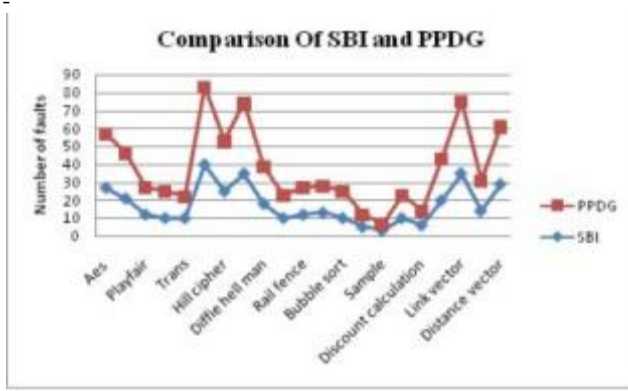

Fig.8 Comparison between SBI and PPDG

The Performance Analysis measure is shown in the fig 6.1. Consider for example a java program Hill Cipher. In this program the number faults identified by SBI is 25. But for the same program PPDG identifies 28 faults. Thus we can confirm that PPDG is efficient than SBI.

## 7. Conclusion and future work

This project proposes an efficient fault localization tool. This tool presents an innovative model for any java program. It scans every internal node and locates where the fault is. PPDG gives a graphical representation of a program. The probabilistic conditional distribution of each node will gives the dependency between the statements in the program. RankCP algorithm ranks the state nodes and the node having less probability will be considered to be most suspicious. In previous work fault localization is done for methods in C programs, PHP, etc. In the proposed work, fault localization is applied for Java programs.

PPDG captures the statistical dependences among program elements and enables the use of probabilistic reasoning to analyze program behaviours. We have used an algorithm for fault localization of the PPDG: RankCP. The result of the study shows the potential usefulness of the PPDG for fault localization. The results also show that the PPDG can be an effective approximate model for representing behaviours of a program for fault diagnosis, eliminating the need to store large amounts of execution information during debugging. Our studies show that, in many cases, RankCP is effective for fault localization. However, the algorithm is not effective in localizing faults in some failing executions. One reason for this ineffectiveness is that RankCP ranks nodes in the PPDG using the conditional probabilities of nodes and their parents. Thus, the algorithm may not localize faults whose effects transcend node-parent state configurations.

We are currently investigating new algorithms that consider local and global effects of faults. Our studies also show that RankCP can be accurate, depending on the context associated with the fault. In practice, it will be beneficial to harness the effectiveness of ranking approaches. One critical part of our PPDG construction is the execution information, which is used to estimate the parameters of the PPDG. In our experiments, we used only passing executions to estimate probability distributions at the nodes in the PPDG. However, we intend to investigate the usefulness of learning PPDG distributions using a combination of passing and failing executions.

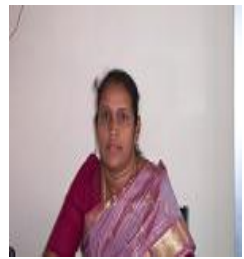

**A.Askarunisa (Dr.)** is currently working as assistant professor in the Department of Computer Science at Thiagarajar College of Engineering , Madurai. She has published number of papers in referred National/ International Journal.

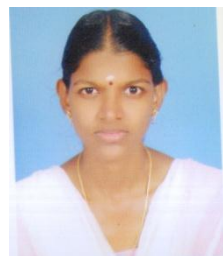

**T. Manju** has received her B.E degree from Vins Christian college of Engineering, Nagercoil affiliated by Anna University, Chennai. Now a post graduate student in the Department of Computer Science at Thiagarajar College of Engineering, Madurai.